# Clarification of the Spontaneous Polarization Direction in Crystals with Wurtzite Structure

Simon Fichtner*,[1,2], Mohamed Yassine[3], Chris van de Walle[4], Oliver Ambacher[3]

[1] Department of Materials Science, Kiel University of California, Santa Barbara, California 93106-5050, USA, Kaiserstr. 2, 24143 Kiel, Germany

[2] Fraunhofer Institute for Silicon Technology, Fraunhoferstr. 1, 25524 Itzehoe, Germany

[3]Institute for Sustainable Systems Engineering (INATECH), University Freiburg, Emmy-Noether-Str. 2, 79110 Freiburg, Germany

[4] Materials Department, University of California, Santa Barbara, California 93106-5050, USA

*sif@tf.uni-kiel.de

**Abstract**

The wurtzite structure is one of the most frequently found crystal structures in modern semiconductors and its inherent spontaneous polarization is a defining materials property. Despite this significance, confusion has been rampant in the literature with respect to the orientation of the spontaneous polarization inside the unit cell of the wurtzite structure, especially for the technologically very relevant III-N compounds (AlN, GaN, InN). In particular, the spontaneous polarization has been reported to either point up or down for the same unit cell orientation, depending on the literature source – with important implications for, e.g., the carrier type and density expected at interfaces of heterostructures involving materials with wurtzite-structure. This perspective aims to resolve this ambiguity by reviewing available reports on the direction of the energetically preferred polarization direction in the presence of external electric fields, as well as atomically resolved scanning transmission electron microscopy images. While we use ferroelectric wurtzite $Al_{1-x}Sc_xN$ as a key example, our conclusions are generalizable to other compounds with the same crystal structure. We demonstrate that a metal-polar unit cell must be associated with an upward polarization vector – which is contrary to long-standing conventional wisdom.

**Introduction and theoretical background**

The wurtzite (*w*) structure [Figure 1] is found in some of the technologically most significant compound semiconductors: AlN, GaN, InN, their ternary alloys, as well as ZnO [1, 2]. With its point group 6mm (dihexagonal-pyramidal), it belongs to the ten spontaneously polarized point groups – unlike the common crystal structures (such as zincblende or diamond) of the other major semiconductor classes. The wurtzite structure belongs to the more symmetric polar structures, with its spontaneous polarization being necessarily parallel to the *c*-axis. This spontaneous polarization is essential to understand not only the properties of materials with *w*-structure, but also the resulting semiconductor devices such as light-emitting diodes (LEDs) and high electron mobility transistors (HEMTs), which have been hugely successful in recent years [3–6].

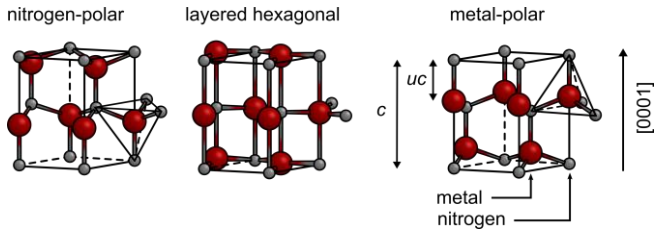

Figure 1: Sketch of the atomic arrangement in the *w*-structure in the case of a III-N compound for both (nitrogen) N- and (metal) M-polarity, as well as the centrosymmetric layered hexagonal structure (which has not been experimentally observed for e.g. AlN, GaN- and InN-based compounds, but serves as a reference structure in computations). The latter is related to the *w*-structure by an increased internal parameter *u* = 0.5. Symmetry dictates that the spontaneous polarization of the *w*-structure is parallel to the *c*-axis of the crystal.

Despite this significance, it has become clear in recent years that the magnitude of the spontaneous polarization $\vec{P}_S$ of materials with *w*-structure had initially been underestimated by more than one order of magnitude. Rigorous theoretical predictions of $\vec{P}_S$ for nitrides were published in 2016 [7], with values > 100 µC/cm² that were far larger than those reported in the seminal calculations by Bernardini, Fiorentini and Vanderbilt in 1997 [2]. Those initial calculations were also methodologically correct; yet a discrepancy arises from the fact that $\vec{P}_S$ always needs to be defined with respect to a *reference structure*. Within the Modern Theory of Polarization, $\vec{P}_S$ is defined as the difference in *formal* polarization between the material of interest and the reference structure [8–10]. In ferroelectrics, where $\vec{P}_S$ can be switched from positive to negative values, the choice of reference structure is intuitive and obvious: the reference is the "symmetric" structure that has zero macroscopic polarization and also has zero *formal polarization*. For *w*-structure materials, the corresponding reference is a layered hexagonal structure [Figure *1*] in which nitrogen and metal atoms reside in the same plane; and this is indeed the reference structure utilized in Ref. [7]. For the purpose of calculating polarization constants, this reference structure is suitable independently of the path and sequence with which individual atoms switch in the real ferroelectric and independently of the fact that it has not been experimentally observed in AlN-, GaN- or InN-based compounds.

In contrast, Ref. [2] chose the zincblende structure as a reference. The problem with this choice is that the formal polarization of zincblende is nonzero along the [111] direction, which is typically aligned with the [0001] direction of the *w*-structure when building supercells for the calculation of the spontaneous polarization [2, 11]. Ignoring this contribution to $P_S$ leads to errors when differences in spontaneous polarization at interfaces are calculated. In spite of this, numerous papers and simulation tools since 1997 have used (and continue to use) the zincblende-referenced $\vec{P}_S$ to predict polarization discontinuities in heterostructures. This fundamental error has gone unnoticed because it is accidentally cancelled by a second error: the use of *proper* rather than *improper* piezoelectric coefficients [12, 13] for calculating the impact of strain on polarization [7], resulting in predictions that closely agree with measurements on experimentally available III-N heterostructures [14].

The choice of reference (layered hexagonal vs. zincblende) affects not only the *magnitude* of $\vec{P}_S$ but also its *direction* relative to the definitions given in Figure *1*. The reason is that the formal polarization in the zincblende structure is slightly larger than the formal polarization in the *w* structure. From calculations, both are large and pointing in the same direction, resulting in $\vec{P}_S$ of the wurtzite structure changing sign when referenced to the zincblende structure. For a metal polar unit cell, one thus obtained a downward facing $\vec{P}_S$. Because of the inadvertent use of the zincblende reference, the notion of a downward pointing $\vec{P}_S$ for metal polar *w*-nitrides has become widespread.

The problems with the choice of the zincblende reference were pointed out in 2016 [7], but did not receive much attention because for the purposes of calculating polarization discontinuities at the interfaces of the readily available III-N heterostructures, the zincblende-referenced values seemed to work well enough (due to the aforementioned cancellation of errors).[14] This changed with the discovery of ferroelectricity in $Al_{1-x}Sc_xN$ in 2019 [15] – a solid solution of AlN and ScN which maintains the *w*-structure up at around 40% ScN. Ferroelectric switching allowed to gain experimental access to $|\vec{P}_S|$ of the *w*-structure for the first time and confirmed the 2016 predictions. The magnitude of $\vec{P}_S$ has since been confirmed independently by more than ten groups for AlN, GaN, and ZnO based compounds [16–18].

While the magnitude of $\vec{P}_S$ can thus be determined with sufficient certainty, significant inconsistency remains in recent publications regarding the direction of the vector $\vec{P}_S$ (which is not directly obvious from e.g. a ferroelectric hysteresis loop). We will use the nomenclature appropriate for *w*-structure III-N semiconductors, where the two polarization directions are named metal (M-) and nitrogen (N-) polar.

The difference in stacking of the nitrogen and metal planes between the two polarities determines the direction of $\vec{P}_S$ and thus the sign of the bound (and compensation) charge on a free surface of a *w*-compound . Confusion abounds in the literature regarding the direction of $\vec{P}_S$. Pre-2016 studies virtually exclusively associate M-polarity with a downward polarization vector, while more recent studies associate M-polarity with either a downward [6, 15, 16, 19–22] or an upward [23–26] polarization vector (relative to Figure *1*; the list of cited publications is far from exhaustive, especially for those reporting the downward direction).

This perspective aims to resolve this inconsistency by referring to published results on ferroelectric switching in $Al_{1-x}Sc_xN$. The assignment of the direction of a vector may seem like a simple task, but it requires the knowledge and connection of two geometrical aspects that are typically not directly observed: the unambiguous direction of the external electric field vector $\vec{E}$ (which depends on the measurement setup) as well as the atomic-scale observation of the unit cell direction. In the following, we will therefore first determine the direction of the electric field vector from depletion effects in n- and p-type substrates and, thus, the energetically preferred direction of the polarization vector under this electric field. Subsequently, we will summarize high-resolution scanning transmission electron microscopy (STEM) results that illustrate the stable unit cell direction under a particular electric field, therefore allowing to connect the polarization direction with the orientation of the unit cell in space. These results unambiguously prove that an M-polar unit cell, as defined in Figure *1*, has an upward-pointing polarization vector – contrary to what can be found in most literature on the subject.

**Electric field and spontaneous polarization direction: ferroelectric switching on p- and n-type substrates**

In this section, we discuss the determination of the direction of the external electric field $\vec{E}$ in a common measurement setup (aixACCT DBLI with drive contact on the upper electrode) through depletion effects during ferroelectric switching of $Al_{1-x}Sc_xN$ on p-doped Si and n-doped GaN, which respectively serve as the bottom electrode during measurement [25, 27]. The electric field $\vec{E}$ direction directly results in the preferred direction of $\vec{P}_S$ (note that $\vec{P}_S$ and $\vec{E}$ are parallel for minimized electric potential energy). This effort is necessary, as the direction of $\vec{E}$ as a consequence of the applied voltage depends on the measurement setup, which also determines the coordinate system used. While theoretical studies and most other sources use a coordinate system [2, 7] with an upward facing z-axis (index 3), piezoelectric $d_{33(,f)}$ coefficient measurements on e.g. sputtered Al(Sc)N are usually reported as positive values. As such films are known to be predominantly N-polar, this implies that these measurements were using a coordinate system where the z-axis is pointing downward [28–30]. In experiment, the choice of coordinate system can be the result of how the sample is contacted, i.e. through which electrode the drive signal is applied. Figure 2 illustrates the difference between the two common choices.

The conclusions drawn in the following with respect to the direction of $\vec{P}_S$ are however fully independent of the choice of the coordinate system/measurement setup. This is directly obvious from Figure 2, as for the external observer, the polarization vector continues to point downward for both setups in case positive external charges are brought to the upper electrode. Care has to be taken only when comparing scalars between the two systems, as their sign might change (e.g. voltage and components of the current or polarization vector, see Table 1).

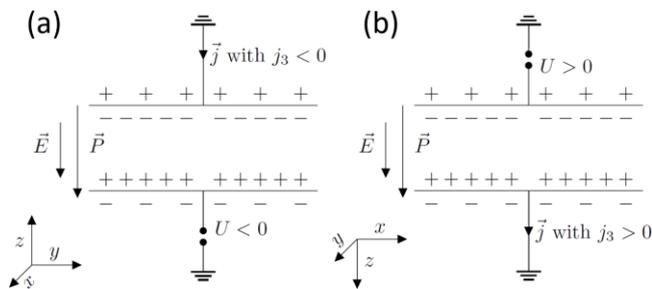

Figure 2: Sketch to illustrate the connection between the directions of $\vec{E}$, $\vec{P}$ and the displacement current density $\vec{j}$ in a ferroelectric as a consequence of a voltage $|U|$ that surpasses the coercive voltage applied either to the bottom (a) or top (b) contact. For clarity, only charge that gives rise to the electric field and polarization charge is sketched, i.e. no screening charge. For the external observer, both situations are identical, but for the sign (not direction!) of the applied voltage and the resulting vectors components.

Based on depletion effects, we will demonstrate which coordinate system was used for the measurements discussed in the following.

During ferroelectric switching, the displacement current density $\vec{J}$ due to the reorientation of $\vec{P}_S$ can be measured and integrated over time to obtain the remanent polarization – which is itself a rough estimate for $|\vec{P}_S|$. This current usually takes the form of a peak centered around the coercive field $E_C$. From Figure 3, it can be observed that the switching peaks are not symmetric for negative and positive electric fields. This behavior can be explained by the formation of a depletion region in the semiconducting substrate, which leads to an additional voltage drop that is less pronounced on symmetric metal electrodes [31]. In the case of Figure 3 (a), it can be observed that ferroelectric switching on a p-doped back electrode shows a reduction in the depletion width and, thus, a thinner switching peak at negative electric fields. In comparison, the formation of a depletion region and, thus, a broader switching peak is present at positive electric fields [27]. At a negative electric field, no depletion occurs during switching, as mobile holes can accumulate in the p-doped semiconductor. A positive external charge at the lower interface implies that negative fixed polarization charges at the lower interface are energetically more favorable (and positive charges at the upper interface). Hence, the polarization direction must be pointing upwards after surpassing the negative coercive field. An even more pronounced depletion occurred on n-doped GaN as the back electrode. As can be observed from Figure 3 (b), the broader peak and thus the formation of a depletion region during the switching process occurs at negative electric fields instead [25]. Positive immobile charges are left at the top interface of the semiconductor back electrode during the polarization reversal under a negative electric field. Consequently, negative polarization charges at the lower interface of the $Al_{1-x}Sc_xN$ are energetically favorable. This observation again confirms that an upward polarization direction results from negative external electric fields in this measurement setup.

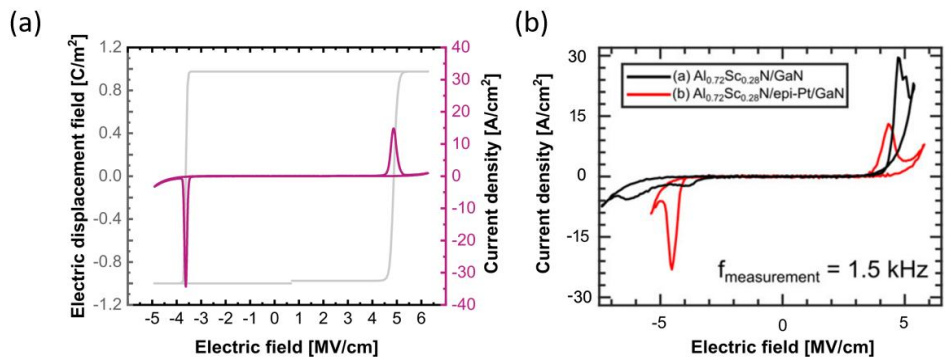

Figure 3: (a) Electric displacement field (approximately equal to polarization) and current density over external electric field for 400 nm thick $Al_{0.7}Sc_{0.3}N$ deposited on p-Si. Depletion is broadening the switching peak on the positive electric field axis. Reproduced from [27],

with the permission of AIP Publishing. (b) Current density over electric field for 100 nm of $Al_{0.72}Sc_{0.28}N$ deposited on (platinized) n-GaN. Depletion significantly broadens the switching peak on the negative electric field axis, in case of directly depositing on the semiconducting interface. Reprinted with permission from *[32]*. Copyright 2023 American Chemical Society.

The direction of $\vec{E}$ was further confirmed through correspondence with aixACCT, the manufacturer of the measurement equipment. Accordingly, connecting the drive head to the upper electrode, the sense head to the lower electrode and establishing a negative electric field results in an upward direction of the latter. Consequently, the more stable polarization direction under these conditions is also upward.

The discussed measurements are therefore represented by Figure 2 (b) (in the following: coordinate system B). Thus, they will result in an upwards pointing polarization vector after surpassing the negative coercive field. For the following discussion, the coordinate system was transformed to the more conventional coordinate system defined by Figure 2 (a) (named coordinate system A). The trivial implications of this coordinate transformation on the quantities used for the following discussion are summarized in Table 1. In coordinate system A, passing the positive coercive field results in an upward facing polarization vector.

Table 1: Summary of the implications of a coordinate transformation between coordinate system A and B for the measurement conditions sketched in Figure 2.

|  | Coordinate system A | Coordinate system B |
|---|---|---|
| $\vec{P}_S, \vec{E}, \vec{j}$ | downward for external observer | downward for external observer |
| $P_3, E_3, j_3, U$ | negative | positive |

**Determining the unit cell orientation from atomically resolved STEM**

Due to the substantial improvement of the texture of ferroelectric thin films with *w*-structure through epitaxial growth, TEM that resolves the local unit cell orientation in a film with *w*-structure has become possible. In the past, convergent beam electron diffraction (CBED) was mainly used for the purpose of determining the unit cell orientation in the III-N compounds [33–35]. Today, the progress of aberration corrected STEM has allowed to also determine unit cell orientation directly from atomically resolved real-space images that reveal the ordering of individual metal and nitrogen planes [23, 32, 36–39]. For our purpose to connect the direction of $\vec{P}_S$ with the orientation of the unit cell, we will refer to two works that use STEM imaging on ferroelectrically switched $Al_{1-x}Sc_xN$ films – i.e. films were the direction of $\vec{P}_S$ can be deduced from electrical measurement data [32, 39]. This allows to identify the unit cell orientation locally, by comparing with their definition given in Figure 1. The connection to the electric field is clearest in ex-situ experiments, by switching individual capacitors from their as-deposited monodomain state with either M- or N-polarity to the opposite polarization prior to sample preparation. The TEM lamella can be prepared such that it includes both an area under the capacitor electrode as well as an area besides it, which has not seen the electric field and thus is still in the initial monodomain states. Alternatively, two TEM lamellas can be prepared which include a switched as well as an unswitched capacitor respectively. This allows to confirm the unit cell orientuation in regions that have been exposed to sufficiently strong electric fields for ferroelectric switching and rules out the formation of domains through the lamella preparation process or the electron irradiation during TEM. Such experimental setups were used in two separate studies [32, 39]. In the first instance, a 550 nm thick Al$_{0.75}$Sc$_{0.25}$N film was deposited by sputter epitaxy on a Mo/Al$_2$O$_3$ template [32]. Using the same measurement setup as discussed above, an upward facing electric field vector with 3.6 MV/cm was established prior to TEM sample preparation. With STEM, regions with M-polarity were found exclusively under the electrode area, while the region outside of the capacitor area was locally confirmed to be N-polar [Figure 4 (a)]. In the second instance, a 230 nm thick Al$_{0.85}$Sc$_{0.15}$N film was deposited on M-polar n-GaN by metal-organic chemical waver deposition (MOCVD) [39]. Unlike the previously investigated sputter-deposited film, the deposition by MOCVD results in single crystal films. This allows the imaging of unit cell orientation over large areas. Prior to the preparation of the TEM sample, the electrical measurement setup was again used to switch a capacitor from its initial monodomain state towards opposite polarity, which required a downward facing electric field vector. STEM imaging was able to confirm that the area outside of the capacitor is exclusively M-polar, while the area under the capacitor could be identified to be predominantly N-polar [Figure 4 (c)]. The reversal of the polarization direction in the unswitched regions of both studies can be explained by the different growth techniques and the impact of the M-polar template on the MOCVD-deposited film. Both studies are nonetheless fully consistent with respect to the connection between the direction of the external electric field and the resulting orientation at the unit cell level. Establishing a downward facing electric field vector led to predominantly N-polar films (starting from originally M-polar films), while an upward facing electric field vector led to the emergence of M-polar regions in previously N-polar films. This is also consistent with results from wet-etching, where upward facing external fields were found to result in slow etching (and thus M-polar) surfaces after ferroelectric switching [15, 25, 26], which can be considered as a more indirect way to connect unit cell orientation and the external electric field [40].

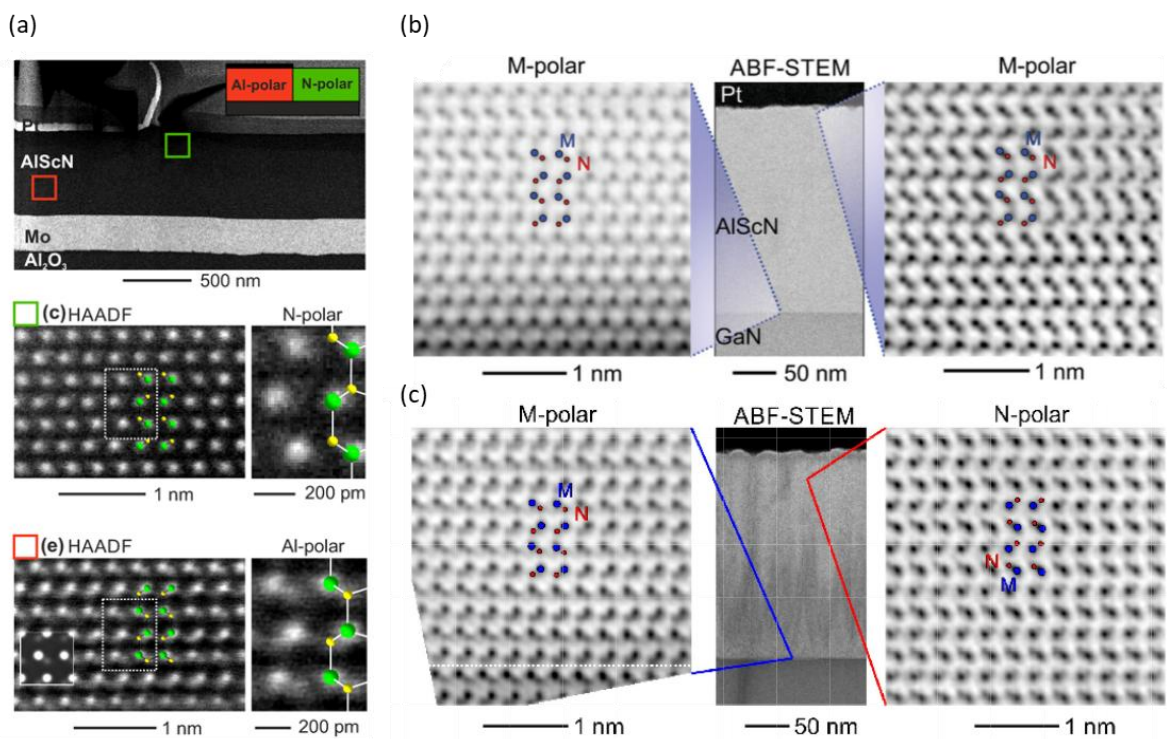

Figure 4: (a) STEM high angle annular dark field images (HAADF) of switched (red) and unswitched (green) regions of a 400 nm thick sputtered Al$_{0.75}$Sc$_{0.25}$N film on Mo/Al$_2$O$_3$.

Switching was performed with an upward facing $\vec{E}$ and resulted in M-polar unit cells. Reproduced from [32] with the permission of AIP Publishing. (b) Annular bright field (ABF) STEM images of an unswitched 200 nm thick $Al_{0.85}Sc_{0.15}N$ film deposited by MOCVD. The film was confirmed to be consistently M-polar after deposition. (c) Same film as in (b) after surpassing the coercive field with a downward facing $\vec{E}$. Most of the film switched to N-polarity. (b) and (c) taken from [39].

**Conclusion**

Based on the observed depletion effects, we confirmed the direction of the electric field under applied bias. From STEM, it was concluded that an upward facing $\vec{E}$ results in an M-polar film (and a downward facing $\vec{E}$ in N-polar films). Therefore, M-polarity must be associated with an upward direction of $\vec{P}_S$ [Figure 5].

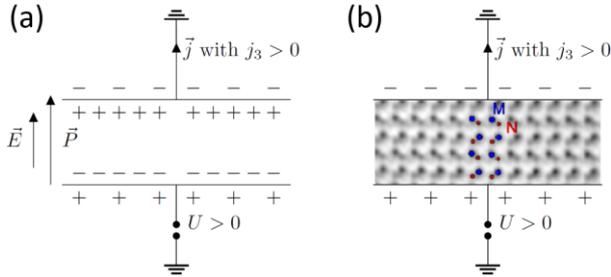

Figure 5: Under the same external biasing, and provided that the resulting $|\vec{E}|$ is larger than the coercive field, an upward pointing $\vec{E}$ and $\vec{P}_S$ can be deduced from depletion effects (panel (a)), while STEM imaging confirms M-polarity (panel (b)). $U$ and $j_3$ are evaluated in coordinate system A.

Furthermore, this upward direction of $\vec{P}_S$ in metal polar films is a necessity to predict the correct carrier type and density in III-N heterostructures such as HEMTs within the correct formulation of the modern theory of polarization, as developed in Ref. [7]. Conversely, the usage of an upward facing $\vec{P}_S$ necessarily requires the usage of the polarization constants referenced to the layered hexagonal structure [7]. This was recently discussed by us in a separate paper on $Al_{1-x}Sc_xN$/GaN heterostructures [26], also considering the piezoelectric polarization[20].

Pre-2016 publications invariably assumed a downward-pointing spontaneous polarization vector in M-polar III-N unit cells, based on the results in Ref. [2]. As explained in the Introduction, a zincblende reference structure was assumed in defining this polarization. Combined with an erroneous usage of proper rather than improper piezoelectric coefficients, the $\vec{P}_S$ values from Ref. [2] resulted in values for polarization discontinuities and bound charges that were consistent with experimentally observations on III-N heterostructures, and hence were unquestioned for almost 20 years. However, as discussed in this Perspective and before, these zincblende-referenced $\vec{P}_S$ values are incapable of describing the actual polarization magnitude in ferroelectric $Al_{1-x}Sc_xN$.[9] Further, as shown in this work, use of $\vec{P}_S$ values referenced to the layered hexagonal structure (which has zero macroscopic polarization) is absolutely essential to correctly describe not only this magnitude (on the order of 100 µC/cm², but in addition also the direction of the polarization vector (pointing upwards as sketched in Figure 5).

An interesting direct consequence of this realization is that polarization discontinuities can be realized in *w* heterostructures that are more than one order of magnitude larger than those based on pre-2016 expectations [20]. Such polarization discontinuities have recently been experimentally shown to be stable, at least in the form of ferroelectric domain walls [17, 24].

We expect that our determination of the direction of $\vec{P}_S$ (when correctly referenced to the layered hexagonal structure with zero polarization) can be generalized to most, if not all, *w*-compounds when associating M-polarity with the more electropositive element and N-polarity with the more electronegative element. We also hope that the present Perspective will contribute to the wider adoption of the $\vec{P}_S$ values referenced to the layered hexagonal structure [7], thereby eliminating the errors and confusion associated with use of spontaneous polarization values referenced to zincblende.

**Conflict of Interest Statement**

The authors have no conflicts to disclose.

**Acknowledgements**

This work was partially supported by the German Science Foundation (DFG) under project no. AM 105/53-1 and FI 2605/1, by the Federal Ministry of Education and Research (BMBF) Ministry under project no. 03VP10842 (VIP+ FeelScreen) and by the Gips-Schüle-Stiftung. Funded by the European Union (ERC, FIXIT, GA 101135398). Views and opinions expressed are however those of the author(s) only and do not necessarily reflect those of the European Union or the European Research Council Executive Agency. Neither the European Union nor the granting authority can be held responsible for them. C. VdW. was supported by SUPREME, one of seven centers in JUMP 2.0, a Semiconductor Research Corporation (SRC) program co-sponsored by DARPA, and by the Army Research Office under grant number W911NF-22-1-0139.